\PassOptionsToPackage{encapsulated}{CJK}
\documentclass{cpbtex}

\makeatletter
\def\@seccntformat#1{%
  \csname the#1\endcsname.%
  \@ifundefined{ccbfmt@#1}{}{\csname ccbfmt@#1\endcsname}%
  \quad}
\@namedef{ccbfmt@section}{\hspace{-.4cm}}
\@namedef{ccbfmt@subsection}{\hspace{-.4cm}}
\@namedef{ccbfmt@subsubsection}{\hspace{-.3cm}}
\makeatother

\usepackage{CJKutf8}
\usepackage{graphicx}
\usepackage{array}
\usepackage{tikz-cd}
\DeclareUnicodeCharacter{2009}{\,}
\DeclareUnicodeCharacter{2212}{\ensuremath{-}}
\AddToHook{env/thebibliography/begin}{%
  \footnotesize
  \setlength{\itemsep}{-3pt plus .2pt minus .2pt}%
}

\begin{document}

\begin{CJK*}{UTF8}{gbsn}

\title{Probing and Manipulating Quantum Materials\\with Strong-field Terahertz and Mid-infrared Radiation}
\author{S. J. Li$^{1,2}$, H. Y. Wu$^{1,2}$, S. J. Zhang$^{1}$ and Y. T. Li$^{1,2,3}$\\[0.5em]
\small $^{1}$Beijing National Laboratory for Condensed Matter Physics, Institute of Physics,\\
\small Chinese Academy of Sciences, Beijing 100190, China\\
\small $^{2}$School of Physical Sciences, University of Chinese Academy of Sciences, Beijing 100049, China\\
\small $^{3}$Attosecond Science Center, Songshan Lake Materials Laboratory, Dongguan, Guangdong 523808, China\\
\small E-mail: sjzh@iphy.ac.cn, ytli@iphy.ac.cn}

\maketitle

\begin{abstract}
Emergent phenomena in quantum materials arise from interactions among different degrees of freedom. Identifying these interactions and establishing their roles in ordered states are central problems in condensed-matter physics and provide a basis for controlling material properties. Many relevant low-energy excitations and gaps lie in the terahertz (THz) and mid-infrared (MIR) range, where intense phase-stable fields can both interrogate nonlinear dynamics and drive selected degrees of freedom far from equilibrium. This review surveys recent progress in probing and manipulating quantum materials with strong-field THz and MIR radiation, with emphasis on nonlinear electronic dynamics, collective-mode coupling, symmetry-sensitive responses and field-induced changes in electronic structure and ordered states. Particular attention is given to developments enabled by the Synergetic Extreme Condition User Facility (SECUF).
\end{abstract}

\section{Introduction}
Charge, spin, orbital and lattice degrees of freedom are simultaneously active and strongly coupled in many quantum materials, and their competition and cooperation can give rise to complex ordered phases\ucite{dagottoComplexityStronglyCorrelated2005}. A broad range of collective excitations and characteristic gaps lie in the THz and MIR range\ucite{basovElectrodynamicsCorrelated2011}. Linear spectroscopy determines their resonance energies, oscillator strengths, linewidths and selection rules, but interactions among different excitations can remain difficult to disentangle in a one-dimensional spectrum\ucite{liuMultidimensionalTerahertzProbes2025}. Nonlinear measurements provide access to harmonic generation, wave mixing, multidimensional correlations and phase-sensitive observables, which can reveal mode coupling, coherent pathways and symmetry constraints\ucite{huang_terahertz_2026}.

Strong THz and MIR fields can couple to electronic motion, lattice vibrations and spin excitations without necessarily creating high-energy interband excitations\ucite{kampfrathResonantNonresonantControl2013a}. Modern sources can reach electric fields in the MV/cm range while providing control over frequency, waveform, phase and polarization\ucite{hiroriSingleCycleTerahertz2011,sellPhaselockedGenerationFieldresolved2008b,satoTerahertzPolarizationPulse2013a}. These capabilities enable resonant excitation, sub-cycle control and access to nonlinear and nonperturbative regimes. At sufficiently high fields, THz and MIR excitation can modify microscopic interactions and drive materials far from equilibrium, providing routes to transiently reshape or transform ordered states\ucite{basovPropertiesDemandQuantum2017,delatorreColloquiumNonthermalPathways2021}.

This review is organized as follows. Section~\ref{sec:para2} summarizes strong-field THz generation, polarization control and phase-sensitive detection. Section~\ref{sec:para3} introduces the principal experimental schemes. Section~\ref{sec:para4} discusses the physical information provided by nonlinear THz spectroscopy. Section~\ref{sec:para5} focuses on experiments in which strong fields modify electronic structure or drive a distinct ordered state. Section~\ref{sec:para6} highlights recent advances enabled by the SECUF platform and discusses future opportunities arising from the integration of strong-field THz excitation with complementary experimental capabilities.

\section{Generation and detection of strong-field THz pulses} \label{sec:para2}

\begin{figure}[b!]
  \centering
  \includegraphics[width=0.82\textwidth]{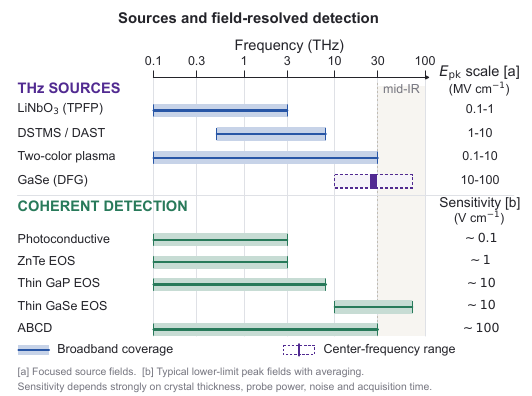}
\caption{Representative frequency coverage of intense laser-driven terahertz sources and coherent detection methods. Solid bands indicate broadband coverage, and dashed boxes denote center-frequency tuning ranges. Representative source parameters are summarized from Refs.~\cite{songReviewIntenseTerahertz2026,mansourzadehIntenseUltrabroadbandHigh2023a,sunTerahertzGenerationLaserinduced2022,liuOpticalTerahertzSources2022}, and detection methods from Refs.~\cite{suterTerahertzTimedomainSpectroscopy2026,wangCoherentDetectionPulsed2025,ahmedDetectivityEnhancementTHz2014,ibrahimUltrahighDynamicRange2016,holmohrtBalancedAirbiasedDetection2024,sellPhaselockedGenerationFieldresolved2008b}.}
  \label{fig:thz-source-detection-coverage}
\end{figure}

Strong-field THz sources include accelerator-based sources and laser-driven tabletop systems. While accelerator sources can provide exceptionally high pulse energies or narrowband tunability\ucite{greenHighFieldTHz2016,helmELBEInfraredTHz2023,kang_continuous_2026}, laser-driven sources are particularly suited to nonlinear spectroscopy because they combine phase stability, synchronization to femtosecond optical pulses, and flexible waveform control\ucite{hoffmannIntenseUltrashortTerahertz2011a}.

Optical rectification is one of the principal routes to intense broadband THz pulses. In LiNbO$_3$, tilted-pulse-front excitation compensates the velocity mismatch between the optical and THz fields and enables efficient generation of high-energy single-cycle pulses\ucite{heblingGenerationHighpower2008,hoffmannIntenseUltrashortTerahertz2011a}. Organic nonlinear crystals such as DAST and DSTMS provide broader spectra extending to higher frequencies, with fields reaching the MV/cm regime\ucite{vicarioGeneration09mJ2014,mansourzadehIntenseUltrabroadbandHigh2023a}. Difference-frequency generation between phase-coherent optical pulses provides spectrally tunable few-cycle or multicycle radiation and can reach extremely high fields in the THz and MIR range\ucite{sellPhaselockedGenerationFieldresolved2008b,Liu2017Generation}. These approaches provide complementary excitation conditions: broadband single-cycle pulses access sub-cycle dynamics, whereas narrowband multicycle pulses enable selective resonant driving.

Laser-induced plasmas provide a complementary crystal-free route to ultrabroadband THz generation. Two-color excitation allows the emitted field to be controlled through the relative phase and polarization of the driving pulses\ucite{xieCoherentControlTHz2006,sunTerahertzGenerationLaserinduced2022}. At substantially higher driving intensities, relativistic laser--foil interactions provide a distinct plasma-based route to extreme THz pulses with peak powers reaching the terawatt regime and spectral tuning controlled by laser and target parameters\ucite{liaoTerawattScale2020,leiHighlyEfficient2022,ruanTerawattTHz2026}. Reviews of this rapidly developing regime can be found in Refs.~\cite{Liao2023Perspectives,Liao2026High}.

Polarization provides an additional degree of freedom for controlling light--matter coupling. Linear polarization selects symmetry-dependent electric- or magnetic-dipole responses, whereas circular and elliptical fields can transfer angular momentum and selectively drive chiral lattice or spin dynamics. Broadband and controllable THz polarization can be generated using polarization optics, coherent field synthesis, or direct polarization control during THz generation\ucite{satoTerahertzPolarizationPulse2013a,jasperCircularTHz2020}.

Quantitative strong-field experiments also require phase-sensitive characterization of the THz waveform. Electro-optic sampling in ZnTe, GaP, GaSe and related crystals remains the most widely used approach\ucite{wuGaPSensor1997,sellPhaselockedGenerationFieldresolved2008b}. Plasma-based coherent detection, including air-biased coherent detection, extends field-resolved measurements over several tens of terahertz\ucite{holmohrtBalancedAirbiasedDetection2024}. Figure~\ref{fig:thz-source-detection-coverage} summarizes representative laser-driven generation and coherent-detection schemes.

Conventional time-domain detection reconstructs the waveform by scanning the delay between the THz field and an optical sampling pulse, whereas single-shot methods encode the temporal profile onto spatial or spectral coordinates\ucite{teoInvitedArticleSingleshot2015}. Such approaches are particularly useful for low-repetition-rate extreme-field sources or irreversible processes, where shot-to-shot reproducibility cannot be assumed\ucite{sunSingleShotTHz2022}.

\section{Experimental schemes for strong-field THz spectroscopy} \label{sec:para3}

Strong-field THz experiments can characterize both the nonlinear response generated during the drive and the state that remains or evolves after excitation\ucite{Dong2023Recent}. High-harmonic generation directly probes nonlinear current or polarization. Multidimensional THz spectroscopy adds one or more independently delayed fields and separates the nonlinear signal as a function of excitation and detection frequencies. Pump--probe measurements characterize the driven state with a weaker delayed probe. THz/MIR-pump--THz-probe spectroscopy measures transient complex conductivity and follows changes in transport, gaps, spectral weight and collective modes. Optical reflection, transmission, polarization, Kerr rotation and SHG provide complementary sensitivity to electronic, magnetic and inversion-symmetry changes. Combined with external tuning parameters, complementary probes such as ultrafast X-ray diffraction, resonant scattering and time-resolved photoemission can relate the driven response to structural, magnetic and electronic changes across the equilibrium phase diagram.

\begin{figure}[t!]
  \centering
  \includegraphics[width=0.96\textwidth]{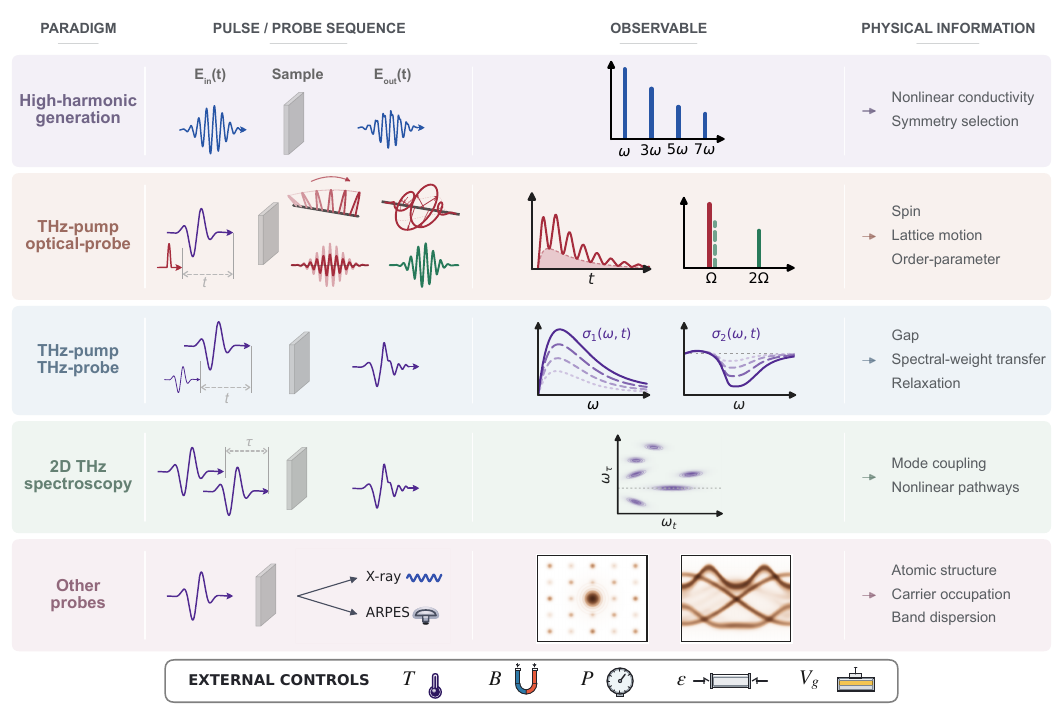}
  \caption{Experimental paradigms for nonlinear THz spectroscopy and control. These representative measurements access nonlinear conductivity, collective-mode and order-parameter dynamics, transient electrodynamics, mode coupling and nonlinear pathways, with external parameters such as temperature, magnetic field, pressure, strain and gate voltage providing additional control.}
  \label{fig:thz-excitation-probe-schemes}
\end{figure}

\section{Physical insights from nonlinear THz spectroscopy} \label{sec:para4}

Strong-field THz spectroscopy provides access to electronic dynamics, coherence, coupling and symmetry beyond the information contained in linear spectra. This section focuses on experiments in which the nonlinear response is used primarily as a spectroscopic observable to identify microscopic dynamics and interactions.

\subsection{Nonlinear electronic dynamics}

\paragraph*{Intraband acceleration}
In the long-wavelength limit, the electromagnetic field can be treated as a spatially uniform force and can shift carriers in momentum space. Within a semiclassical single-band picture, the carrier dynamics are governed by the acceleration of the crystal momentum, $\hbar \dot{\mathbf{k}}=-e\mathbf{E}(t)$. The resulting current is determined by how the carrier velocity evolves along the band dispersion. For a parabolic band with a momentum-independent relaxation rate, the velocity remains proportional to momentum and the resulting current is approximately linear in the field. 

Band-structure-induced nonlinear current emerges when the field drives carriers into regions of nonparabolic or periodic band dispersion. For example, in n-doped GaAs, intense single-cycle THz fields displaced electrons over a substantial fraction of the Brillouin zone. Field-resolved emission revealed coherent ballistic motion approaching a partial Bloch oscillation, directly connecting the nonlinear waveform to the periodic band dispersion\ucite{kuehnCoherentBallisticMotion2010}. In InGaAs, transfer into satellite valleys with lower mobility abruptly modified the current on a sub-cycle timescale and generated high-frequency radiation\ucite{chaiSubcycleTerahertzNonlinear2018}.

Scattering can also produce a nonlinear response. In the Hund metal CaRuO$_3$, enhanced third-harmonic generation appears together with the formation of coherent
heavy quasiparticles below approximately $80~\mathrm{K}$, with energy-dependent scattering identified as an important source of the nonlinearity\ucite{reinhofferStrongTerahertzThirdharmonic2024a}. Measurements across the Mott insulator--metal transition in rare-earth nickelates further show that THz harmonics are sensitive to changes in the microscopic transport channels and scattering dynamics\ucite{prajapatiTerahertzHarmonic2026}. 

A particular example of nonlinear intraband dynamics is provided by Dirac bands. For an ideal linear dispersion, the velocity magnitude is approximately fixed while its direction follows the driven momentum. A large oscillatory momentum shift therefore produces a nonsinusoidal current, motivating early predictions of efficient frequency multiplication in graphene\ucite{mikhailovNonlinearElectromagnetic2007,mikhailovNonlinearElectromagnetic2008}. Experiments indeed observed harmonics up to seventh order in graphene, where THz heating and the resulting modulation of intraband conductivity played an important role\ucite{hafezExtremelyEfficientTerahertz2018a}. In the three-dimensional Dirac semimetal Cd$_3$As$_2$, polarization-resolved measurements and Boltzmann calculations support a regime dominated by coherent intraband acceleration and an anisotropic nonequilibrium distribution\ucite{chengEfficientTerahertzHarmonic2020c,kovalevNonperturbativeTerahertzHighharmonic2020a}. 

\paragraph*{Interband excitation and electron--hole dynamics}

When the drive accesses interband transitions through resonant or multiphoton excitation, or becomes strong enough to induce interband tunneling, the response is no longer restricted to carriers already present within a band. Interband excitation creates electron--hole pairs, which can subsequently be accelerated by the same field. In n-type GaAs, intense THz transients induced interband tunneling, with its strong temperature dependence revealing the importance of ultrafast interband decoherence\ucite{kuehnTerahertzinducedInterbandTunneling2010}. In GaSe, phase-stable high harmonics extending from the THz to the visible arose from the combined action of intraband Bloch motion and interband polarization, with waveform-dependent interference between distinct excitation pathways\ucite{schubertSubcycleControlTerahertz2014d}.

The electron--hole motion can also produce particularly direct signatures in high-order sideband generation. A near-infrared pulse creates an electron--hole pair, after which the carriers are separated and accelerated by a strong THz field before recombination generates a ladder of optical sidebands\ucite{zaksExperimentalObservationElectronhole}. Polarization-resolved sideband measurements in bulk GaAs subsequently exploited quantum interference between different recollision pathways to reconstruct hole Bloch wavefunctions\ucite{costelloReconstructionBloch2021}. More recently, time- and spectrum-resolved measurements of THz-driven excitonic Autler--Townes states in MoS$_2$ revealed a sharp increase in dephasing above the field threshold for exciton dissociation, together with nonperturbative high-order sideband generation below this threshold\ucite{liuDephasingStrongfielddrivenExcitonic2024a}.

\subsection{Nonlinear spectroscopy of collective modes}

Nonlinear THz spectroscopy provides access to collective-mode dynamics, interactions and hybridization beyond the information available from linear spectra.

\paragraph*{Superconducting amplitude dynamics}

The superconducting order parameter can be written as $\Delta=|\Delta|e^{i\phi}$. The Higgs mode corresponds to oscillations of $|\Delta|$ and couples only weakly to electromagnetic fields in linear response, making nonlinear THz excitation a principal route for its detection\ucite{shimanoHiggsModeSuperconductors2020}. The first direct THz observation of superconducting amplitude dynamics was reported in Nb$_{1-x}$Ti$_x$N, where a single-cycle pulse launched a freely evolving Higgs oscillation at the gap scale\ucite{matsunagaHiggsAmplitudeMode2013}. Subsequent experiments in NbN established resonantly enhanced THG and higher-order harmonic responses associated with driven amplitude dynamics\ucite{matsunagaLightinducedCollectivePseudospin2014c,wangTransientHiggsOscillations2022}. 

Nevertheless, several microscopic channels can produce nonlinear responses on the same energy and temperature scales. Quasiparticles, charge-density fluctuations and Cooper-pair fluctuations may overlap spectrally with the amplitude-mode response, while retardation, multiband effects and impurity scattering modify their relative contributions\ucite{ceaNonlinearOpticalEffects2016,tsujiNonlinearLightHiggs2016,tsujiHiggsModeResonance2020}. In multiband or intertwined systems, nonlinear THz spectroscopy revealed band-selective amplitude responses and dynamical interference between superconducting and charge-density-wave amplitude modes\ucite{kovalevBandSelectiveThirdHarmonic2021,yuanSelectiveExcitation2025,fengDynamicalInterplaySuperconductivity2023b}. Nonlinear THz measurements in cuprates have likewise been used to follow superconducting-order dynamics and their coupling to competing electronic orders\ucite{kimTracingDynamicsSuperconducting2024,katsumiHiggsCuprate2018,chuFano2023,yuanPseudogapTHG2024}.

More recently, multidimensional THz spectroscopy has begun to separate these microscopic pathways rather than simply identify nonlinear resonances. Experiments have resolved coherent Higgs--quasiparticle interactions in Nb and disorder-activated amplitude responses in NbN\ucite{huangHiggsEcho2025,katsumiRevealingNovelAspects2024}, while magnetic-field measurements of strongly disordered NbN showed that a substantial normal-state THG channel can coexist with the superconducting response\ucite{wangDisorderedNbN2026}. These results illustrate why frequency-, phase- and field-dependent measurements are essential for distinguishing amplitude-mode dynamics from competing quasiparticle and collective contributions.

\paragraph*{Interlayer coherence in cuprates}
\label{par:josephson_spectroscopy}

In layered cuprates, coherent tunneling between adjacent CuO$_2$ planes gives rise to the Josephson plasma resonance. Its intrinsic nonlinearity follows from $j_s=j_c\sin\phi$, where $\phi$ is the gauge-invariant interlayer phase difference. Intense THz excitation has provided direct access to nonlinear phase dynamics in addition to the linear plasma frequency\ucite{rajasekaranParametricAmplificationSuperconducting2016a}. A Josephson plasma echo in two-dimensional spectroscopy separated homogeneous dephasing of the interlayer phase from a static distribution of local plasma frequencies\ucite{liuProbingInhomogeneousCuprate2024a}. Nevertheless, the nonlinear signals near the plasma edge require careful interpretation. THz-pump--THz-probe measurements showed that the apparent oscillation frequency can vary systematically with probe frequency because Josephson nonlinearity and the frequency-dependent electrodynamic response of the sample jointly shape the detected waveform\ucite{zhangJosephsonNSR2023}.

Nonlinear Josephson electrodynamics can also reveal interlayer pair coherence that is weak or silent in the linear optical response. In stripe-ordered La$_{1.885}$Ba$_{0.115}$CuO$_4$, a giant THz third harmonic persisted above $T_c$ up to the charge-ordering temperature and was interpreted within a pair-density-wave picture as nonlinear mixing of frustrated interlayer tunneling modes\ucite{rajasekaranProbingOpticallySilent2018a}. Driven lattice modes provide another route to nonlinear Josephson dynamics. In YBa$_2$Cu$_3$O$_{6+x}$, resonant phonon excitation produced signals consistent with parametric amplification of Josephson plasma polaritons, and multidimensional MIR spectroscopy subsequently resolved the underlying phonon-mediated coupling\ucite{vonhoegenAmplificationSuperconducting2022,taherianProbingAmplifiedJosephson2025}.

\paragraph*{Anharmonic lattice dynamics and mode coupling}

The harmonic normal-mode picture becomes incomplete once lattice displacements are large enough for higher-order terms in the potential-energy surface to become relevant. The lattice potential may be expanded as
\begin{equation}
V(\{Q_i\})=
\frac{1}{2}\sum_i\Omega_i^2Q_i^2
+\sum_{ijk}g_{ijk}Q_iQ_jQ_k
+\sum_{ijkl}h_{ijkl}Q_iQ_jQ_kQ_l+\cdots ,
\end{equation}
where the quadratic term defines independent harmonic normal modes, while the higher-order terms describe both the anharmonicity of individual modes and nonlinear coupling between different lattice coordinates. 
Single-mode terms such as $Q_i^3$ and $Q_i^4$ distort the oscillatory trajectory, leading to harmonic generation and amplitude-dependent frequency shifts. 
In LiNbO$_3$, for example, the strongly driven $A_1$ phonon was described by a potential containing both cubic and quartic contributions, and harmonics up to fifth order were observed and used to reconstruct the sampled interatomic potential\ucite{vonhoegenProbingInteratomicPotential2018e}. 
In SrTiO$_3$, quartic anharmonicity of the THz-driven soft mode produced amplitude-dependent hardening and saturation, while higher-order intermode coupling transferred energy to higher-frequency optical phonons\ucite{kozinaTerahertzdrivenPhononUpconversion2019b}. Anharmonicity also couples otherwise independent normal modes. Mixed terms such as $Q_a^2Q_b$ generate forces on one coordinate that depend on the motion of another, providing the microscopic basis for rectification, frequency conversion, and nonlinear phononics\ucite{forstNonlinearPhononicsUltrafast2011,subediTheoryNonlinearPhononics2014}. 

Multidimensional THz spectroscopy can further resolve the nonlinear pathways. 
For example, in LiNbO$_3$, 2D THz--THz--Raman spectroscopy showed that two different phonons are excited predominantly through resonant one-photon THz absorption, while the observed nonlinear cross-peaks are most consistent with mechanical anharmonicity of the lattice potential rather than nonlinearities of the dipole or polarizability operators\ucite{linMappingLiNbO32022}. 

\paragraph*{Excitation pathways of Raman-active phonons}

In a centrosymmetric crystal, Raman-active modes have even parity and therefore do not couple linearly to the electric field through a dipole term proportional to \(E Q_{\mathrm R}\). Several pathways can be represented schematically by the driving forces
\begin{equation}
F_{\mathrm R}^{\mathrm{phot}}
\propto R_{ij}E_iE_j,\qquad
F_{\mathrm R}^{\mathrm{phon}}
\propto gQ_{\mathrm{IR}}^2,\qquad
F_{\mathrm R}^{\mathrm{mix}}
\propto \lambda E Q_{\mathrm{IR}},
\end{equation}
where $R_{ij}\propto\partial\chi_{ij}/\partial Q_{\mathrm R}$ is the Raman tensor and $Q_{\mathrm{IR}}$ denotes a coherently driven infrared-active phonon.

The first term describes a polarizability-mediated photonic Raman process. 
Because the product $E_iE_j$ contains both difference- and sum-frequency components, a Raman mode can be driven when either frequency combination matches its eigenfrequency. 
A particularly clear example is diamond, where the infrared-inactive $F_{2g}$ optical phonon at approximately $40~\mathrm{THz}$ was coherently excited by the sum of two frequency components of an intense THz pulse centered near half the phonon frequency\ucite{maehrleinTerahertzSumFrequency2017}. 
The second pathway is phonon mediated: an infrared-active mode is first driven linearly by the THz field, and anharmonic lattice coupling such as $Q_{\mathrm{IR}}^2Q_{\mathrm R}$ subsequently exerts a force on the Raman coordinate. 
This ionic-Raman mechanism underlies nonlinear phononics and can generate either a rectified displacement or coherent oscillation of the Raman mode, depending on the mode frequencies and coupling\ucite{forstNonlinearPhononicsUltrafast2011,subediTheoryNonlinearPhononics2014,juraschekSumfrequencyIonicRaman2018}. 
The third term represents a mixed photophononic pathway, in which the electromagnetic field and a coherently driven infrared phonon act together on the Raman coordinate through a coupling proportional to $E Q_{\mathrm{IR}}Q_{\mathrm R}$.

Field scaling alone is generally insufficient to distinguish these pathways, because different mechanisms can produce similar low-order nonlinear responses. Two-dimensional THz spectroscopy provides a more direct separation through their distinct excitation--detection frequency correlations. 
In CdWO$_4$, it distinguished electronic Raman excitation from anharmonic phonon-mediated driving\ucite{johnsonDistinguishingNonlinear2019}, whereas in MnBi$_2$Te$_4$ it identified a mixed photophononic pathway in which the THz field first prepares an infrared-active phonon and subsequently couples to this coherent motion to excite a Raman-active mode\ucite{blankTwoDimensionalTerahertz2023}. Propagation effects provide an additional complication, for phase matching between the THz pump and optical probe can generate oscillatory features that do not correspond to intrinsic collective modes\ucite{shenKerrPolarization2026}. 
Reliable pathway assignment therefore requires consistency among frequency correlations, crystal symmetry, field dependence, and propagation geometry.

\paragraph*{Magnon coherence and nonlinear coupling}

For magnetic excitations, higher harmonics, wave mixing and echoes constrain nonlinear terms in the spin Hamiltonian and correlations between magnon branches. Two-dimensional THz spectroscopy of YFeO$_3$ resolved rephasing spin echoes and two-quantum signals, providing access to coherent dephasing and correlations involving pairs of magnon excitations\ucite{luCoherentTwodimensionalTerahertz2017a}. Sum- and difference-frequency generation between distinct magnon branches subsequently established nonlinear magnon--magnon coupling in the same material and provided polarization selection rules for the interaction\ucite{zhangTerahertzFieldinducedNonlinear2024b}.

Further measurements extend this picture from mode coupling to high-order spin dynamics. In ErFeO$_3$, nonlinear interactions enabled upconversion from a lower- to a higher-frequency magnon branch\ucite{zhangMagnonUpconversion2024}. In Sm$_{0.4}$Er$_{0.6}$FeO$_3$, harmonics up to sixth order and seventh-order wave mixing constrained the roles of magnetic anisotropy and Dzyaloshinskii--Moriya interactions through comparison with spin simulations\ucite{huangExtremeTerahertzMagnon2024c}. In FePS$_3$, double-THz-pump spectroscopy resolved magnon--phonon sum- and difference-frequency signals, directly identifying coherent three-wave mixing between spin and lattice excitations\ucite{liMagnonPhononMixing2026}. The appearance of interaction-generated frequencies is particularly valuable because it distinguishes coupled dynamics from the eigenfrequencies already visible in linear spectra.

\paragraph*{THz-driven polariton dynamics}

Strong THz pulses can launch propagating hybrid modes when the electromagnetic field couples resonantly to infrared-active lattice excitations. A phonon-polariton is characterized by a frequency- and wavevector-dependent mixture of electromagnetic and lattice components, so its propagation provides access to dispersion, group velocity, and damping in addition to the bare phonon frequency. 
In NiI$_2$ and MnPS$_3$, broadband THz pulses launched phonon-polariton wave packets that were tracked by time-resolved second-harmonic generation. 
Time-of-flight and attenuation measurements reconstructed the real and imaginary parts of the polariton dispersion, while an anomalous feature in NiI$_2$ revealed a hidden magnon--phonon interaction\ucite{luoTimeFlightPhononPolariton2024}. 
In ZnO, high-field THz excitation generated propagating phonon-polaritons whose repeated passage through a millimeter-scale crystal modulated the optical second-harmonic response for tens of picoseconds\ucite{fangCoherentManipulationPhononPolariton2025}. 
These experiments extend strong-field THz spectroscopy from local collective-coordinate dynamics to propagating hybrid excitations, where the relevant observables include not only mode frequency and amplitude but also dispersion, attenuation, and transport.

\subsection{Symmetry-sensitive nonlinear THz spectroscopy}

Nonlinear optical susceptibilities are intrinsically sensitive to crystal symmetry. Topological insulators provide a direct example of the distinction between surface and bulk symmetry. In Bi$_2$Se$_3$, THz second-harmonic generation was nearly independent of film thickness and disappeared in topologically trivial controls, consistent with a nonlinear current dominated by Dirac-like surface states because electric-dipole SHG is forbidden in the centrosymmetric bulk\ucite{stensbergObservationTerahertzSecond2024a}. In a complementary experiment, resonant excitation of an infrared-active $E_u$ phonon produced large time-dependent changes in optical SHG\ucite{bowlanProbingControlling2017}. The response comprised a fundamental component at the phonon frequency from the inversion-broken surface and a second-harmonic component from coherent phonon motion in the bulk, enabling spectral separation of surface and bulk symmetry dynamics within the same material.

In metallic Pb$_{1-x}$Sn$_x$Te, resonant THz excitation combined with time-resolved SHG exposed soft-mode dynamics and a distortive polar order that is difficult to identify in linear spectroscopy because of the strong Drude background\ucite{shiRevealingDistortivePolar2024b}. A related use of THz-induced SHG in the quantum paraelectric KTaO$_3$ revealed long-lived polar correlations following soft-mode excitation\ucite{chengTerahertzdrivenLocalDipolar2023c}. In ZrTe$_5$, a coherently driven infrared-active phonon acted as a local oscillator for an extremely weak equilibrium $\chi^{(2)}$. The heterodyne signal, together with its field scaling and polarization dependence, revealed pre-existing bulk inversion-symmetry breaking that conventional SHG did not resolve\ucite{liZrTe5Heterodyne2026}. 

\section{Strong-field control of electronic structure and ordered states} \label{sec:para5}

Apart from diagnosing microscopic interactions, strong THz/MIR fields can reshape electronic structure, cross free-energy barriers or stabilize transient and metastable ordered states.

\subsection{Control of electronic structure and electronic phases}

\paragraph*{Coherent reshaping of electronic structure}

Strong THz and MIR fields can reshape the electronic Hamiltonian before substantial scattering and thermalization occur. Two useful limiting pictures are the Stark effect, in which the instantaneous field shifts and mixes electronic levels, and Floquet dressing, in which a periodic drive hybridizes states separated by integer multiples of the photon energy.

The Stark picture is especially intuitive when the field acts primarily as a time-dependent electric bias. In quantum-confined systems, an out-of-plane field can modify the electron and hole wave functions and shift excitonic transitions through the quantum-confined Stark effect (QCSE)\ucite{millerBandEdgeElectroabsorption1984}. Strong THz pulses bring this field control into the sub-picosecond regime. In few-layer MoS$_2$, for example, a THz nanoantenna generated an enhanced out-of-plane field
approaching the MV/cm scale, causing the A- and B-exciton resonances to follow the THz waveform on a sub-cycle timescale\ucite{hiraokaTerahertzFieldEffect2025a}. 

For the Floquet picture, a periodic drive can generate photon-dressed replica bands and open gaps where these replicas hybridize. Time- and angle-resolved photoemission first provided direct evidence of Floquet--Bloch states on the surface of a topological
insulator Bi$_2$Se$_3$\ucite{wangObservationFloquetBloch2013}. Floquet engineering was subsequently demonstrated in a semiconductor by a near-resonant MIR excitation, which produced strong band renormalization, Floquet sidebands and light-induced gaps\ucite{zhouPseudospinSelectiveFloquet2023}. Below-gap excitation further showed that momentum-dependent band renormalization can persist even without resonant excitation across the equilibrium band gap\ucite{zhouFloquetEngineeringBlack2023a}.

\paragraph*{Field-induced electronic breakdown and phase transitions}

Strong THz fields can drive electronic breakdown even when the photon energy lies far below the electronic gap. In contrast to conventional photoexcitation, sufficiently strong low-frequency fields can generate real charge excitations through field-induced tunneling rather than direct photon absorption. In Mott insulators, this process can be described as a many-body Landau--Zener or Schwinger-like breakdown, in which doublon--holon pairs are created across the correlation gap\ucite{okaDielectricBreakdownMott2010,ecksteinDielectricBreakdownMott2010}. 

This mechanism has been observed in several correlated systems. In \(\kappa\)-(ET)$_2$Cu[N(CN)$_2$]Br, a strong THz field with photon energy far below the Mott gap generated doublon--holon pairs through field-induced tunneling and induced collapse of the gap within approximately $0.1~\mathrm{ps}$\ucite{yamakawaMottTransitionImpulsive2017c}. V$_2$O$_3$ further illustrates the competition between electronic and thermal pathways: far below the equilibrium transition, the response is dominated by prompt field-induced tunneling, whereas closer to the phase boundary enhanced Joule heating promotes slower metallic-domain growth\ucite{giorgianniOvercomingThermalRegime2019}. Similar peak-field-driven metallization has been reported in VO$_2$\ucite{mayerTunnelingBreakdownStrongly2015c}, while in the candidate excitonic insulator Ta$_2$NiSe$_5$ intense THz fields induce tunneling-assisted exciton dissociation and partial melting of the excitonic phase\ucite{takamuraMeltingExcitonicInsulator2024}. 

\subsection{Resonant collective-mode control of ordered states}

Resonant excitation provides a mode-selective route for controlling ordered
states because the electromagnetic field can preferentially drive a selected collective coordinate. The connection between the driven mode and the target state can be organized into three useful pathways as discussed below.

\paragraph*{Direct driving of functional collective modes}

The most direct situation occurs when an optically active collective mode is itself a principal coordinate of the ordered state. Resonant excitation can then generate a large-amplitude displacement or rotation along the relevant energy landscape without requiring an intermediate mode. A canonical example is the soft polar mode of quantum-paraelectric SrTiO$_3$. Intense THz excitation drives this mode to large amplitude and produces a threshold-like second-harmonic response together with changes in the low-frequency phonon spectrum, consistent with the formation of a transient polar state\ucite{liTerahertzFieldInduced2019}. Theoretical modeling further suggested how the driven soft-mode coordinate, lattice expansion and dissipation can cooperate to stabilize a displaced configuration\ucite{shinSimulatingTerahertzFieldinduced2022b}. 

Direct mode control is not restricted to lattice polarization. In multiferroic TbMnO$_3$, an intense THz electric field resonant with an electromagnon drove a coherent rotation of the spin cycloid, providing a particularly clear example in which an electric-dipole-active collective spin excitation gives direct access to magnetic order\ucite{kubackaLargeAmplitudeSpin2014}. Likewise, the low-frequency collective dynamics of polar skyrmions in PbTiO$_3$/SrTiO$_3$ superlattices can couple directly to THz fields. Resonant excitation of these collective polar modes drives the skyrmion texture toward a hidden state with a transient macroscopic polarization\ucite{liPolarSkyrmions2025}. 

\paragraph*{Indirect control through auxiliary-mode coupling}

When the target coordinate is weakly infrared active, dipole-forbidden, or located outside the accessible spectral range, an alternative strategy is to resonantly excite an auxiliary mode and use nonlinear coupling to transfer the drive to another lattice, electronic or magnetic coordinate. This principle underlies nonlinear phononics, in which an infrared-active phonon can generate a directional displacement or coherent motion of another mode through lattice anharmonicity\ucite{forstNonlinearPhononicsUltrafast2011,subediTheoryNonlinearPhononics2014}.

One of the clearest realizations is LiNbO$_3$, where the ferroelectric mode is not driven directly. Resonant excitation of a higher-frequency infrared-active phonon instead couples anharmonically to the ferroelectric coordinate and transiently displaces the lattice across the centrosymmetric configuration, producing a reversal of polarization\ucite{mankowskyUltrafastReversalFerroelectric2017a}. In centrosymmetric PbTe, resonant THz driving of a soft transverse-optical coordinate activates strong anharmonic coupling to the orthogonal degenerate TO component, producing a transient symmetry-broken ferroelectric state\ucite{huangCoherentSeedingFerroelectricity2026}. A related symmetry-based route was demonstrated in CoF$_2$. Two driven infrared phonons couple through three-phonon mixing to a Raman-active distortion that makes the two antiferromagnetic sublattices inequivalent and thereby generates a transient ferrimagnetic moment\ucite{disaPolarizingAntiferromagnetOptical2020}. In
nonchiral BPO$_4$, excitation of selected components of a pair of degenerate phonons similarly couples to symmetry-lowering lattice coordinates and creates a transient chiral structure\ucite{zengPhotoinducedChiralityNonchiral2025b}. 

Auxiliary phonons can also act on magnetic order by modifying microscopic
interactions rather than by simply displacing another normal mode. In
DyFeO$_3$, resonant MIR excitation of an infrared-active phonon modifies the
Fe--Dy interaction and thereby reshapes the magnetic free-energy landscape,
enabling switching between competing magnetic configurations near the
spin-reorientation transition\ucite{afanasievUltrafastControlMagnetic2021a}. In YTiO$_3$, different resonantly driven phonons modify the orbital polarization and the proximity to competing magnetic phases, producing opposite changes in the magnetic response depending on the selected mode\ucite{disaPhotoinducedHightemperatureFerromagnetism2023}. In FePS$_3$, THz
excitation was likewise associated with phonon-induced changes in exchange
couplings, demonstrating how a driven lattice coordinate can reorganize a
magnetic energy landscape\ucite{ilyasMetastableMagnetization2024}.

The same indirect-control principle extends beyond phonon-mediated pathways. In Sm$_{0.7}$Er$_{0.3}$FeO$_3$, resonant excitation of the q-AFM mode couples nonlinearly to the slower q-FM coordinate, reshaping the magnetic energy landscape and enabling switching near the spin-reorientation transition\ucite{zhangSpinSwitchingSm07Er03FeO32025a}. This experiment provides
a magnetic analogue of auxiliary-mode control.

Resonantly driven phonons have also been extensively explored as an indirect
route to superconducting correlations\ucite{buzziPhotomolecular2020}. In
underdoped YBa$_2$Cu$_3$O$_{6+x}$, MIR excitation of an infrared-active
$B_{1u}$ apical-oxygen phonon couples anharmonically to fully symmetric
$A_g$ Raman coordinates and reorganizes the lattice structure
\ucite{mankowskyNonlinearLatticeDynamics2014,mankowskyCoherentModulation2015};
under related conditions, transient low-frequency electrodynamics resembling
enhanced interlayer coherence and, more recently, a diamagnetic response have
been reported\ucite{huOpticallyEnhancedCoherent2014,favaMagneticExpulsion2024}.
The interpretation of these responses remains debated up to now, for other
measurements in cuprates have emphasized photoexcited quasiparticles and the
coexistence of a $c$-axis reflectivity edge with enhanced in-plane scattering
\ucite{zhangPhotoinducedNonequilibrium2020,zhangLightInducedMelting2024,Dong2023Recent}.
Likewise, MIR excitation of K$_3$C$_{60}$ has produced gap-like optical and
nonlinear transport responses\ucite{mitranoPossibleLightInduced2016,roweResonantSuperconductivity2023},
whereas excitation-depth effects, optical saturation and the absence of a
resolved magnetic-expulsion signal place important constraints on their
interpretation\ucite{dodgeOpticalSaturation2023,devecchiExpulsionLimit2025}.

\paragraph*{Helicity-selective control through phonon angular momentum}

A qualitatively different control parameter emerges when two orthogonal
phonon coordinates are driven with a well-defined relative phase. Circular or
elliptical lattice motion carries angular momentum whose direction reverses
with the helicity of the excitation. Instead of producing only a scalar
displacement of the lattice or modifying an interaction strength, the driven phonon can then generate an axial effective field that couples to magnetic or ferroic degrees of freedom\ucite{juraschekChiralPhonons2025}.

An early demonstration was provided by ErFeO$_3$, where elliptically driven
lattice motion generated an effective magnetic field and launched coherent
spin precession\ucite{novaEffectiveMagneticField2017}. In CeF$_3$,
circularly driven phonons produced an effective magnetic field of order
$1~\mathrm{T}$ and transiently polarized paramagnetic spins
\ucite{luoChiralPhonons2023}. In SrTiO$_3$, circular motion of a degenerate
soft phonon generated a room-temperature magnetization through dynamical
multiferroicity\ucite{basiniTerahertzElectricfielddrivenDynamical2024b}.

The same principle can be extended from transient responses to deterministic
switching. Circular phonons generated in a paramagnetic substrate have been
used to reverse the magnetization of an adjacent GdFeCo layer through the
ultrafast Barnett effect, with the switching direction selected by the pump
helicity\ucite{daviesPhononicSwitching2024}. In the ferroaxial material
RbFe(MoO$_4$)$_2$, circular excitation of a degenerate infrared phonon
produces a helicity-dependent axial conjugate field that selects between
opposite ferroaxial domains, enabling nonvolatile and rewritable switching
\ucite{zengPhotoinducedNonvolatileRewritable}. These results illustrate a
distinct form of mode-selective control in which the handedness of the driven
motion, rather than only its amplitude, determines the direction of the
induced response.

\section{Advances and perspectives in strong-field THz science at SECUF} \label{sec:para6}

SECUF provides a distinctive combination of spectrally flexible strong-field excitation, multidimensional and phase-sensitive THz detection, and precise control of external thermodynamic parameters. For example, the A4 THz/MIR station enables strong-field excitation under low temperature and high magnetic field conditions, while the dual-source platform allows independent control of the spectrum, polarization, intensity and delay of two strong THz pulses\ucite{wangSECUFSpectroscopy2026,wangTabletopLaserbasedTerahertz2024a,wangVersatileTwodimensionalTerahertz2026}. The D3 laser--solid source further extends the accessible field regime toward terawatt-level THz peak power with tunable center frequencies from $3$ to $20~\mathrm{THz}$\ucite{ruanTerawattTHz2026}. 

These capabilities have already enabled studies that move strong-field THz spectroscopy beyond the observation of nonlinear responses toward understanding microscopic interactions and controlling emergent states. Representative examples include separating competing nonlinear channels in disordered NbN through magnetic-field tuning, resolving coherent magnon--phonon coupling in FePS$_3$ using multidimensional THz spectroscopy, revealing hidden equilibrium inversion-symmetry breaking in ZrTe$_5$ through phonon-assisted heterodyne amplification, and creating transient polar states in PbTe and polar-skyrmion systems through strong-field phonon control\ucite{wangDisorderedNbN2026,liMagnonPhononMixing2026,liZrTe5Heterodyne2026,huangCoherentSeedingFerroelectricity2026,liPolarSkyrmions2025}.

Future advances in strong-field THz science at SECUF will benefit from deeper integration of THz/MIR excitation with the diverse experimental capabilities already available at the facility. Combining strong-field excitation with time-resolved photoemission, ultrafast electron diffraction, ultrafast X-ray techniques and other momentum-, structure- and symmetry-sensitive measurements will enable the coupled evolution of electronic, lattice and magnetic degrees of freedom to be directly tracked within the same driven-state framework. Such an integrated approach will provide new opportunities to establish connections between microscopic interactions and emergent nonequilibrium states, revealing how driven quantum phases arise and evolve under external excitation.

\begin{center}
\small{\textbf{ACKNOWLEDGMENTS}}
\end{center}
This work was supported by the National Key Research and Development Program of China (No.~2024YFA1408703) and the National Natural Science Foundation of China (Nos.~U24A2016, 12304184 and 12574164). 

\bibliographystyle{unsrt}
\addcontentsline{toc}{section}{References}
\bibliography{review}

\end{CJK*}
\end{document}